%% file: tp_pap.tex
\documentclass[twocolumn]{jpsj2} %% two-column layout
\title{Thermopower of Multiorbital Kondo Effect via Single Quantum Dot System at Finite Temperatures}

\author{Rui \textsc{Sakano}$^{1}$\thanks{E-mail address: sakano@tp.ap.eng.osaka-u.ac.jp}, Tomoko \textsc{Kita}$^{1}$ and Norio \textsc{Kawakami}$^{1,2}$}

\inst{$^{1}$Department of Applied Physics, Osaka University, Suita, Osaka 565-0871, Japan \\
$^{2}$Department of Physics, Kyoto University, Kyoto 606-8502, Japan}

\abst{We study the thermopower and some related transport quantities due to the orbital Kondo effect in a single quantum dot system with a finite value of Coulomb repulsion by means of the noncrossing approximation applied to the multiorbital impurity Anderson model. It is elucidated  how the asymmetry of the renormalized tunneling resonance due to the two-orbital Kondo effect causes characteristic behavior of the thermopower at finite temperatures under gate-voltage and magnetic-field control, which is compared with that of the ordinary spin Kondo effect.}

\kword{thermopower, figure of merit, quantum dot, Kondo effect, orbital degrees of freedom}

\begin{document}
\maketitle

\input{Sec/intro.tex}
\input{Sec/m_and_c.tex}
\input{Sec/results.tex}

\input{Sec/summary.tex}

\input{Sec/acknowledgment.tex}

%\appendix
%\input{Sec/to_nca.tex}

%%%%%%%%%%%%%%%%%%%%%%%%%%%%%%%%%%%%%%%%%%%%%%%%%%%%
%%%%%%%%%%%%%%%%%%%%%%%%%%%%%%%%%%%%%%%%%%%%%%%%%%%%

\end{document}

%% file: Sec/intro.tex
\section{Introduction} %% No sections necessary for express letters, letters and short notes

The Kondo effect due to magnetic impurity scattering in metals is a well known and widely studied phenomenon
\cite{pap:Kondo,book:hewson}.
The effect has recently received much renewed attention since it was found that the Kondo effect significantly influences the conductance in quantum dot (QD) systems at low temperatures
\cite{pap:D.GG}.
A lot of tunable parameters in QD systems have made it possible to systematically investigate electron correlations.
In particular, we focus on the orbital properties of QDs, which are induced by high symmetry in shape of a single QD or multiplicity of coupled QDs.
Experimentally, they were observed in vertical QDs, carbon nanotube QDs, self-assemble QDs and multiple QDs systems,
whose orbital levels and splittings are controlled by gate voltage and magnetic fields.
\cite{pap:tarucha1,pap:white,pap:Cobden,pap:PJH_cnt1,pap:Moriyama,pap:PJH_cnt2,pap:Bellucci}
This has stimulated extensive studies on the conductance due to the orbital Kondo effect;
the singlet-triplet Kondo effect
\cite{pap:sasaki_st,pap:st_Eto,pap:st_Pustilnik1,pap:st_KKikoin,pap:Izumida,pap:imai,pap:st_Hofstetter,pap:st_Pustilnik2},
doublet-doublet or \textit{SU}(4) Kondo effect
\cite{pap:Hettler,pap:Yeyati,pap:Sasaki2,pap:pjh,pap:Choi,pap:sakano,pap:Hur,pap:Busser1,Sakano2}
and the orbital Kondo effect in multiple QD systems
\cite{pap:Wilhelm,pap:Sun,pap:Borda,pap:sakanoDD,pap:Galpin,pap:Lipinski,pap:Mravlje,pap:Nishikawa,pap:Kuzmenko1}.

The thermopower we study in this paper is another important transport quantity, which gives complementary information on the density of states to the conductance measurement:
the thermopower can sensitively probe the asymmetric nature of the tunneling resonance around the Fermi level.
So far, a few theoretical studies have been done on the thermopower in QD systems
\cite{pap:Beenakker,pap:boese,pap:Turek,pap:tskim,pap:Matveev,pap:BDong,pap:Krawiec,pap:Donabidowicz}. Remarkably, the thermopower due to the spin Kondo effect  was observed in a recent experiment on a lateral QD system
\cite{pap:Scheibner},
which naturally motivates us to theoretically explore this transport quantity in more detail.

Here, we discuss how the asymmetry of the renormalized tunneling resonance due to the orbital Kondo effect affects the thermopower under gate-voltage and magnetic-field control. By employing the noncrossing approximation (NCA) for the Anderson model with finite Coulomb repulsion, we especially investigate the Kondo effect in several electron-charge regions of the QD. We also discuss the thermoelectric figure of merit, which indicates applicability of our multiorbital QD system to thermopower generators.
%%%%%%%%%%%%%%%%%%%%%%%%%%%%%%
%In this connection, it was pointed out that orbital degeneracy induces large va%lues of thermopower in strongly correlated systems of transition metal oxides
%\cite{pap:Koshibae}.
%%%%%%%%%%%%%%%%%%
This is partly motivated by a recent theoretical assertion in a slightly different context that orbital degeneracy induces large values of thermopower in strongly correlated systems of transition metal oxides
\cite{pap:Koshibae}.

This paper is organized as follows.
In Sec. \ref{sec:m_and_c}, we introduce the generalized Anderson impurity model for our QD with orbital degrees of freedom, and outline how to treat transport properties by the mean of the NCA. In Sec. \ref{sec:results}, we first
discuss the thermopower and some related transport quantities due to the ordinary spin Kondo effect, and then present the results for the two-orbital Kondo effect as an example of generic multiorbital Kondo effects. We clarify how the asymmetric Kondo resonance formed in the presence of the orbital degrees of freedom controls the sign and the strength of the thermopower. Finally, a brief summary is given in Sec. \ref{sec:summary}.

%% file: Sec/m_and_c.tex
%%%%%%%%%%%%%%%%%%%%%%%%%%%%%%%%%%%%%%%%%%%%%%%%%%%%%
\section{Model and calculation}\label{sec:m_and_c}%%%
%%%%%%%%%%%%%%%%%%%%%%%%%%%%%%%%%%%%%%%%%%%%%%%%%%%%%

%%%%%%%%%%%%%%%%%%%%%%%%%%%%%%%%%%
\subsection{Quantum dot system with orbitals}%%%
\begin{figure}[tb]
\begin{center}
\includegraphics[width=0.5\linewidth]{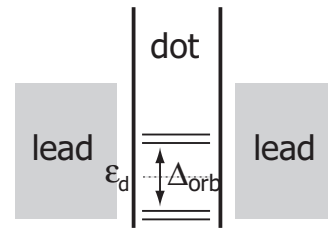}
\end{center}
\caption{
Energy-level scheme of a single QD system with orbital degrees of freedom coupled to two leads.
Applied magnetic fields give rise to orbital splittings $\Delta_{orb}$.
}
\label{fig:sch}
\end{figure}

%%%%%%%%%%%%%%%%%%%%%%%%%%%%%%%%%%
Let us consider a single QD system with degenerate orbitals in equilibrium, as shown in Fig. \ref{fig:sch}.
%%%%%%%%%%%%%%%%%%%%%
%%%%%%%%%%%%%%%%%%%%%
The energy levels of the QD are assumed to be,
%%%%%%%%%%%%%%%%%%%%%
\begin{eqnarray}
&&\varepsilon_{\sigma l} = \varepsilon_d + l \Delta_{orb}, \\
&&l=-(N_{orb}-1)/2,-(N_{orb}-3)/2, \cdots, (N_{orb}-1)/2 \nonumber
\end{eqnarray}
%%%%%%%%%%%%%%%%%%%%%
where $\varepsilon_d$ denotes the center of the energy levels, $\sigma$ ($l$) is a spin (orbital) index and $N_{orb}$ represents the degree of orbital degeneracy.
In practice, this type of orbital splitting has been experimentally realized as Fock-Darwin states in vertical QD systems or clockwise and counterclockwise states in carbon nanotube QD systems.
In these QD systems, the energy-level splitting between the orbitals $\Delta_{orb}$ is induced in the presence of magnetic field $B$ or deformation of QDs; the orbital splitting is usually proportional to the magnetic field $\Delta_{orb} \propto B$, so that the orbital splittings are experimentally tunable
\cite{pap:tarucha1,pap:Tokura}.
In addition, the Zeeman splitting is assumed to be much smaller than the orbital splitting, so that we can ignore the Zeeman effect.

Our QD system is described by the multiorbital Anderson impurity model,
%%%%%%%%%%%%%%%%%%%%%
\begin{eqnarray}
{\cal H} &=& {\cal H}_l + {\cal H}_d + {\cal H}_{t} \label{eq:hamiltonian} \\
%%%%%%%%
{\cal H}_l &=& \sum_{k \sigma l} \varepsilon_{k \sigma l} c^{\dagger}_{k \sigma l} c_{k \sigma l}, \\
%%%%%%%%%%
{\cal H}_d &=& \sum_{k \sigma l} \varepsilon_{\sigma l} d^{\dagger}_{\sigma l} d_{\sigma l}
+ U \sum_{\sigma l \neq \sigma' l'} n_{\sigma l} n_{\sigma' l'} \nonumber \\
&& \qquad -J \sum_{l \neq l'} \mbox{\boldmath$S$}_{dl} \cdot \mbox{\boldmath$S$}_{dl'} , \\
%%%%%%%%%
{\cal H}_{t} &=&  \sum_{k \sigma } V_{k \sigma l} \left( c^{\dagger}_{k \sigma l} d_{\sigma l} + \mbox{H. c.} \right),
\end{eqnarray}
%%%%%%%%%%%%%%%%%%%%%%
where $c^{\dagger}_{k\sigma l} (c_{k\sigma l})$ creates (annihilates) a conduction electron with wave number $k$, spin $\sigma=\pm 1/2$ and orbital $l$
and, similarly, $d^{\dagger}_{\sigma l} (d_{\sigma l})$ creates (annihilates) an electron in the QD.
Here, $U$ is the Coulomb repulsion and $J(>0)$ represents the Hund coupling in the QD. In this Hamiltonian, we have introduced the orbital indices for the conduction electrons. Since orbital and spin moments are conserved in tunneling of vertical QD or carbon nanotube QD systems, this assumption may be justified for these systems
\cite{pap:sasaki_st,pap:Sasaki2,pap:pjh}.

The non-equilibrium Green function technique allows us to study various transport properties, which gives the expressions for the electric conductance, the thermopower and the thermal conductance as
\cite{pap:BDong},
%%%%%%%%%%%%%%%%%%%%%%%
\begin{eqnarray}
G &=& -\frac{e^2}{T}{\cal L}_{11} \\
S &=& -\frac{1}{eT} \frac{{\cal L}_{12}}{{\cal L}_{11}}, \\
\kappa &=& \frac{1}{T^2}\left( {\cal L}_{22} - \frac{{{\cal L}_{12}}^2}{{\cal L}_{11}} \right),
\end{eqnarray}
%%%%%%%%%%%%%%%%%%%%%%%%
with the linear-response coefficients,
%%%%%%%%%%%%%%%%%%%%%
\begin{eqnarray}
{\cal L}_{11} &=& \frac{\pi T}{h} \Gamma \sum_{\sigma l} \int d\varepsilon \, \rho_{\sigma l}(\varepsilon) \left( - \frac{\partial f(\varepsilon)}{\partial \varepsilon} \right), \label{eq:Lrc1} \\
{\cal L}_{12} &=& \frac{\pi T}{h} \Gamma \sum_{\sigma l} \int d\varepsilon \, \varepsilon \rho_{\sigma l} (\varepsilon) \left( - \frac{\partial f(\varepsilon)}{\partial \varepsilon} \right), \label{eq:Lrc2} \\
{\cal L}_{22} &=& \frac{\pi T}{h} \Gamma \sum_{\sigma l} \int d\varepsilon \, \varepsilon^2 \rho_{\sigma l} (\varepsilon) \left( - \frac{\partial f(\varepsilon)}{\partial \varepsilon} \right). \label{eq:Lrc3}
\end{eqnarray}
%%%%%%%%%%%%%%%%%%%%%
%%%%%%%%%%%
Here, we have assumed that resonance widths due to $V_{k \sigma l}$ between the QD and the left (right) lead $\Gamma_L$ ($\Gamma_R$) are equal to $\Gamma$, for simplicity. In general, we have to treat  them independently
with the relation $\Gamma=2\Gamma_L\Gamma_R/[\Gamma_L+\Gamma_R]$. Although
the asymmetry between $\Gamma_L$  and $\Gamma_R$ may change
the magnitude of the transport quantities, we expect that it does not change 
their characteristic properties qualitatively.
Note that it also alters the Kondo temperature, but we can 
incorporate the resulting effects by properly rescaling the physical 
quantities by the Kondo temperature
\cite{pap:Choi,pap:Coqblin}.

%%%%%%%%%%%%%%%%%%%%%%%%%%%%%%%%%%%%%%%%%
\subsection{Noncrossing approximation}%%%
%%%%%%%%%%%%%%%%%%%%%%%%%%%%%%%%%%%%%%%%%
The NCA is a selfconsistent perturbation theory, which summarizes a specific series of expansions in the hybridization $V$.
This method is known to give physically sensible results at temperatures near or higher than the Kondo temperature
\cite{pap:Bickers}.
In fact, it was successfully applied to the Ce and Yb impurity problem
\cite{pap:Kuramoto,pap:Zhang,pap:Coleman,pap:Maekawa,pap:Pruschke}
or used as an impurity solver for dynamical mean field theory, for which orbital degrees of freedom play an important role
\cite{pap:Kotoliar}.

The NCA basic equations can be obtained in terms of coupled equations for the self-energies $\Sigma_m(z)$ of the resolvents $R_m(z)=1/[z-\varepsilon_m - \Sigma_m(z)]$,
%%%%%%%%%%%%%%%%%%%
\begin{eqnarray}
\Sigma_m(z) &=& \frac{\Gamma}{\pi} \sum_{m'} \sum_{\sigma l} \left[ \left( M^{\sigma l}_{m' m} \right)^2 + \left( M^{\sigma l}_{m m'} \right)^2 \right] \nonumber \\
&& \qquad \times \int d\varepsilon R_{m'}(z+\varepsilon)f(\varepsilon),
\end{eqnarray}
%%%%%%%%%%%%%%%%%%%
where the index $m$ specifies the eigenstates of ${\cal H}_d$ 
and the mixing width is $\Gamma=\pi \rho_c V^2$ with the density of states of conduction electrons $\rho_c$.
The coefficients $M_{mm'}^{\sigma l}$ are determined by the expansion coefficients of the fermion operator,
%%%%%%%%%%%%%%%%%%%
\begin{eqnarray}
d_{\sigma l}^{\dagger}=\sum_{mm'} M_{mm'}^{\sigma l} | m \rangle \langle m' | .
\end{eqnarray}
%%%%%%%%%%%%%%%%%%%
The local density of states in the QD is given by,
%%%%%%%%%%%%%%%%%%%
\begin{eqnarray}
 \rho_{\sigma l}(\varepsilon) = \frac{1}{Z} \sum_{mm'} |M^{\sigma l}_{mm'}|^2 \int d\varepsilon' \, e^{-\beta \varepsilon} \qquad \qquad \qquad && \nonumber \\
 \times \left[ \rho_m(\varepsilon') \rho_{m'}(\varepsilon' + \varepsilon) + \rho_m(\varepsilon' -\varepsilon) \rho_m'(\varepsilon) \right] &&
\end{eqnarray}
%%%%%%%%%%%%%%%%%%%
with the partition function,
\begin{eqnarray}
Z= \sum_m \int d\varepsilon \, e^{-\beta \varepsilon} \rho_m (\varepsilon),
\end{eqnarray}
%%%%%%%%%%%%%%%%%%%
and the spectral function of the resolvent $R_m(\varepsilon)$,
%%%%%%%%%%%%%%%%%%%
\begin{eqnarray}
\rho_m(\varepsilon) = - \frac{1}{\pi}\mbox{Im}R_m(\varepsilon+i\delta).
\end{eqnarray}
%%%%%%%%%%%%%%%%%%%
This completes our formulation of the transport coefficients within the 
NCA method.  By evaluating the density of states $\rho_m(\varepsilon)$ numerically, we can investigate the thermopower and the related transport quantities.

%% file: Sec/results.tex
%%%%%%%%%%%%%%%%%%%%%%%%%%%%%%%%%%%%%%%%%%%%%%%%%
\section{Numerical results}\label{sec:results}%%%
%%%%%%%%%%%%%%%%%%%%%%%%%%%%%%%%%%%%%%%%%%%%%%%%%

%%%%%%%%%%%%%%%%%%%%%%%%%%%%%%%%%%%%%%%%%%%%%
\subsection{Spin Kondo effect; $N_{orb}=1$}%%
%%%%%%%%%%%%%%%%%%%%%%%%%%%%%%%%%%%%%%%%%%%%%

\begin{figure*}[tb]
\begin{center}
\includegraphics[width=0.8\linewidth]{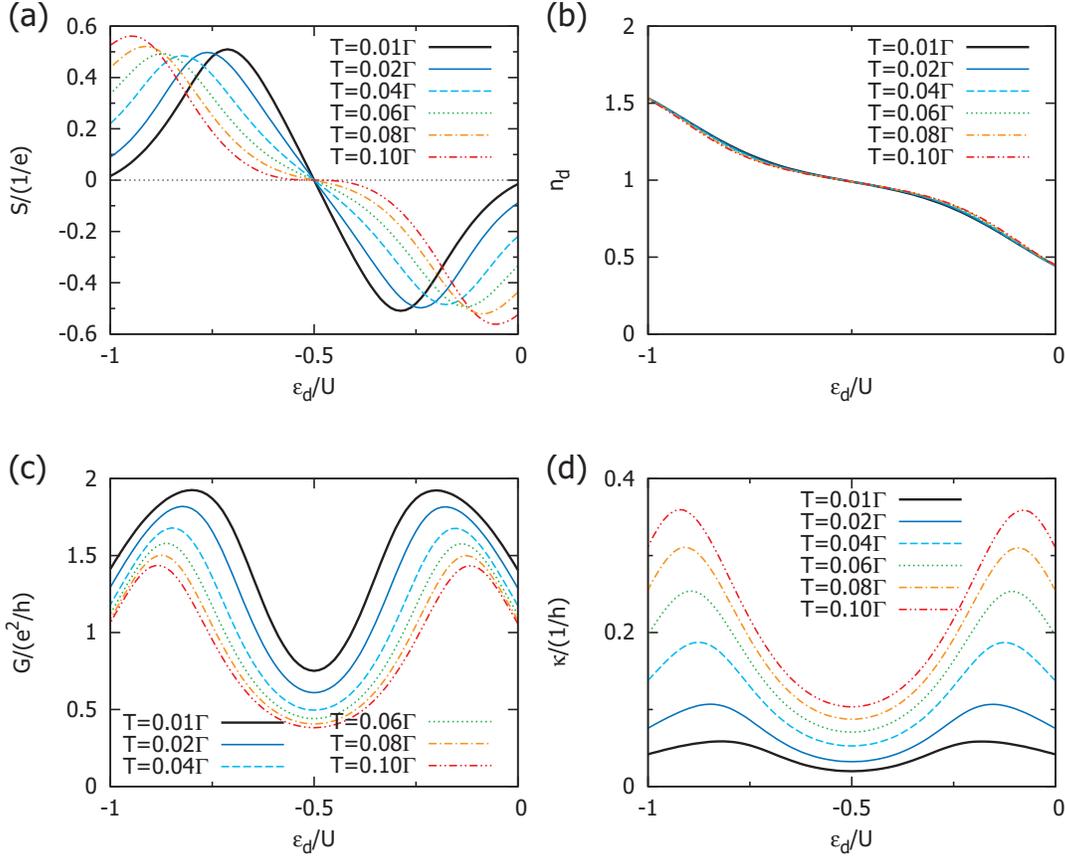}
\end{center}
\caption{(Color online) Transport quantities due to the spin Kondo effect ($N_{orb}=1$): (a)thermopower, (b)number of electrons in the QD, (c)conductance, and (d)thermal conductance, as a function of the QD level (or equivalently gate voltage).
We set $U=8\Gamma$.}
\label{fig:sorbtp}
\end{figure*}

\begin{figure*}[tb]
\begin{center}
\includegraphics[width=0.8\linewidth]{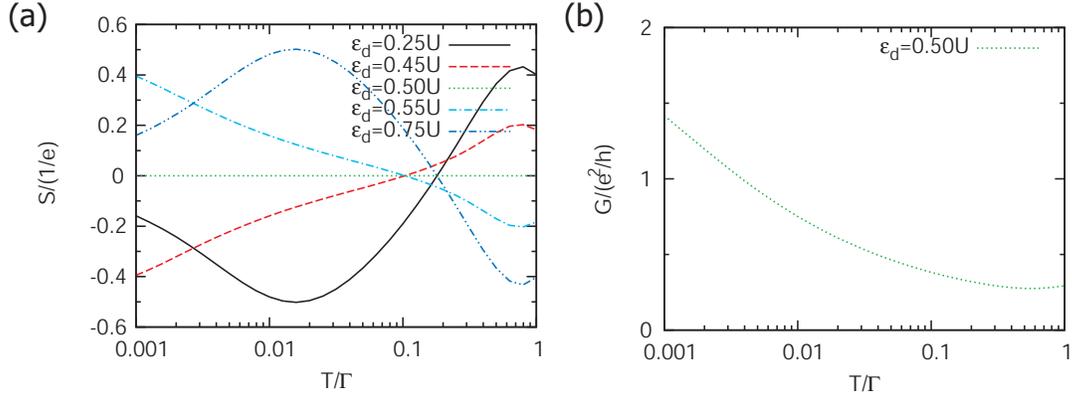}
\end{center}
\caption{(Color online) Transport quantities due to the spin Kondo effect ($N_{orb}=1$): (a)temperature dependence of the thermopower for several chaoices pf QD levels around $\varepsilon_d=-U/2$, (b)temperature dependence of the conductance for $\varepsilon_d=-U/2$.
We set $U=8\Gamma$.}
\label{fig:sorbtd}
\end{figure*}

Let us start with  transport properties due to the ordinary 
spin Kondo effect ($N_{orb}=1$).
The computed linear-response transport coefficients
 as a function of the gate-voltage ({\it i.e.} QD levels $\varepsilon_d$) are shown in Fig. \ref{fig:sorbtp}.
%%%%%%%%%%%%%%%%%%%%%
%%%%%%%%%%%%%%%%%%%%%
The conductance exhibits two Coulomb peaks near $\varepsilon_d/U \sim 0,-1$ at higher temperatures in Fig. \ref{fig:sorbtp}(c).
As well known,  when the temperature decreases, the ordinary {\it SU}(2) Kondo effect due to the spin degrees of freedom becomes dominant and thus increases the conductance in the region $-1<\varepsilon_d/U < 0$ with $n_d \sim 1$ \cite{pap:Ng2,pap:Glazman1,pap:izumida2,pap:Gerace}.
Such enhancement of the spin Kondo effect is indeed observed in the conductance shown in Fig. \ref{fig:sorbtp}(c). On the other hand, the thermopower in Fig. \ref{fig:sorbtp}(a) turns out to vanish around $\varepsilon_d/U=-1/2$ while its first derivative becomes very large at low temperatures.
This suggests that the sharp Kondo resonance is indeed formed at low temperatures, and its peak position is located just at the  Fermi level for $\varepsilon_d/U=-1/2$. Therefore, when the dot level is slightly changed from this condition, the position of the Kondo resonance is shifted across the Fermi level, which causes the sign change of the thermopower.
The thermal conductance shown in  Fig. \ref{fig:sorbtp}(d) is not sensitive to the detail of the Kondo resonance in comparison with the thermopower.
We will use this quantity to discuss the figure of merit of the system (see below).
For reference, we show the temperature dependence of the thermopower and the
conductance in  Fig. \ref{fig:sorbtd}.
In contrast to monotonic behavior in the conductance, the thermopower drastically changes its character around $\varepsilon_d/U=-1/2$,
%%%%%%%%%%%%%%%%%%%%%%
%%%%%%%%%%%%%%%%%%%%%%
which confirms that even small perturbations could easily change the sign of the thermopower at low temperatures.

The above characteristic properties of the thermopower due to the {\it spin} Kondo effect have been experimentally observed in a lateral QD system
\cite{pap:Scheibner},
and theoretically discussed by means of the equation of motion method with finite Coulomb repulsion
\cite{pap:Donabidowicz}.
Our results discussed in this section agree with these observations.

\begin{figure}[tb]
\begin{center}
\includegraphics[width=0.8\linewidth]{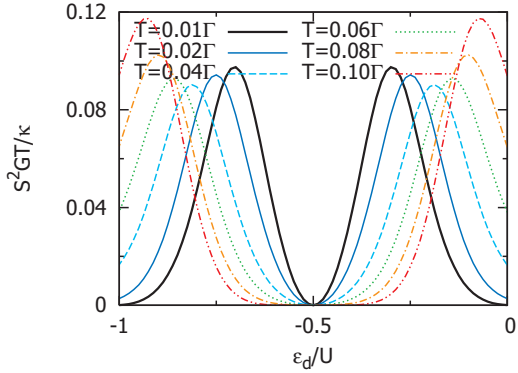}
\end{center}
\caption{(Color online) Figure of merit $ZT$ due to the spin Kondo effect ($N_{orb}=1$), for several choices of temperatures, as a function of QD level.
We set $U=8\Gamma$.}
\label{fig:s_vFOM}
\end{figure}

Before closing this subsection, we briefly comment on applicability of the QD system for thermoelectric devices.
Thermoelectric figure of merit $Z=S^2G/\kappa$ is a measure of the usefulness of materials or devices for thermopower generators or cooling systems.
Since the thermoelectric figure of merit for simple systems is inversely proportional to the temperature $T$, it is convenient to plot $ZT$, which indicates the system performance.
In Fig. \ref{fig:s_vFOM}, we show $ZT=S^2GT/\kappa$ at several temperatures as a function of the QD level.
%%%%%%%%%%%%%%%%%%%%%%
%%%%%%%%%%%%%%%%%%%%%%
We find that the values are much smaller than unity in the whole parameter regime.
This means that the practical applicability is quite limited 
near the Kondo temperature or higher temperatures, as far as the spin Kondo effect is concerned. We will see below that the situation is improved substantially
if we take into account the orbital degrees of freedom.

%%%%%%%%%%%%%%%%%%%%%%%%%%%%%%%%%%%%%%%%%%%%%%%%%%%%%
\subsection{Two-Orbital Kondo effect; $N_{orb}=2$}%%%
%%%%%%%%%%%%%%%%%%%%%%%%%%%%%%%%%%%%%%%%%%%%%%%%%%%%%

Let us now turn to  the \textit{orbital} Kondo effect in the QD system.
We discuss the gate-voltage and magnetic-field dependence of the transport 
quantities by taking the two-orbital Kondo effect as an example.

%%%%%%%%%%%%%%%%%%%%%%%%%%%%%%%%%%%%%%%
\subsubsection{Gate-voltage control}%%%
\begin{figure*}[tb]
\begin{center}
\includegraphics[width=0.8\linewidth]{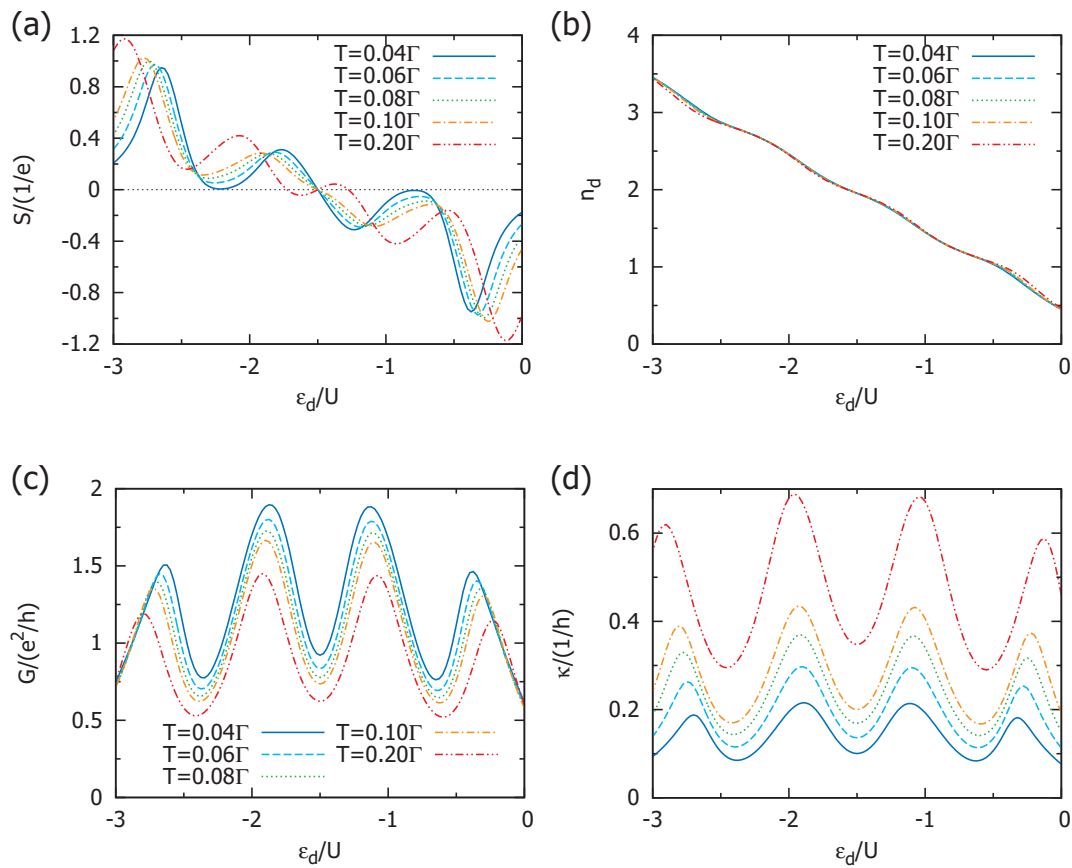}
\end{center}
\caption{(Color online) Transport quantities due to the two-orbital Kondo effect ($N_{orb}=2$):(a)thermopower, (b)number of electrons in the QD, (c)conductance, and (d)thermal conductance, as a function of the QD level.
We set $J=0$ and $U=8\Gamma$.}
\label{fig:2orbtp}
\end{figure*}

\begin{figure*}[tb]
\begin{center}
\includegraphics[width=0.8\linewidth]{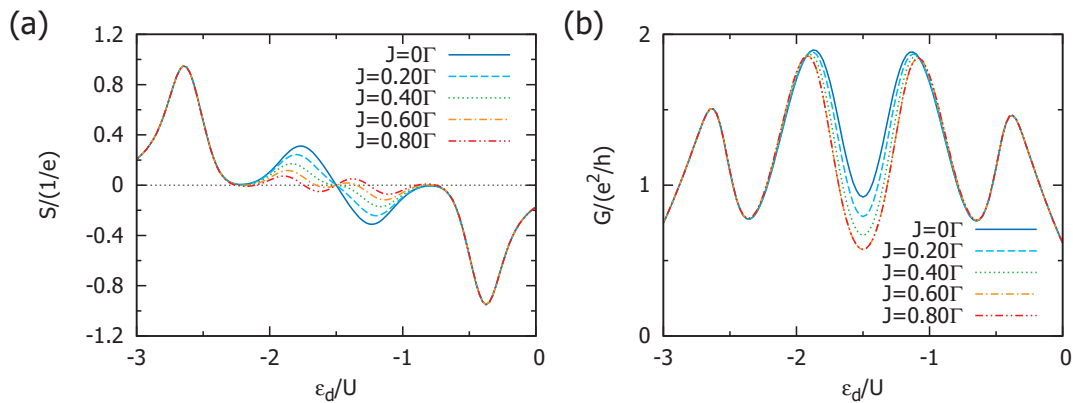}
\end{center}
\caption{(Color online) (a)Thermopower and (b)conductance, due to the two-orbital Kondo effect ($N_{orb}=2$), as a function of the QD level.
We set $T=0.04\Gamma$ and $U=8\Gamma$.}
\label{fig:double_vjSG}
\end{figure*}

%%%%%%%%%%%%%%%%%%%%%%%%%%%%%%%%%%%%%%%
The computed transport quantities for the two-orbital Kondo effect ($N_{orb}=2$)  are shown in Fig. \ref{fig:2orbtp}, where we assume the Hund coupling $J=0$.
%%%%%%%%%%%%%%%%%%%%%
%%%%%%%%%%%%%%%%%%%%%
We first notice that there are four Coulomb peaks around $\varepsilon_d/U \sim 0,-1,-2,-3$  in the conductance shown in Fig. \ref{fig:2orbtp}(c) at high temperatures.  This is also the case for the thermal conductance (Fig. \ref{fig:2orbtp}(d)). However, as seen from Fig. \ref{fig:2orbtp}(a), the thermopower does not show such simple sawtooth behavior in the gate-voltage dependence, which is in contrast to the spin Kondo effect discussed above \cite{pap:Scheibner,pap:Donabidowicz}.
As the temperature decreases, the thermopower in the region of $-1<\varepsilon_d/U < 0 (-3<\varepsilon_d/U < -2)$ with $n_d \sim 1 (3)$ is dominated  by the four-component Kondo effect (so-called {\it SU}(4) Kondo effect) that includes spin and orbital degrees of freedom. The thermopower has negative values in the region $-1<\varepsilon_d/U < 0$,
implying that the effective tunneling resonance due to the Kondo effect is located slightly above the Fermi level.
At low enough temperatures, the {\it SU}(4) Kondo effect is enhanced 
until $\varepsilon_d $ is lowered down to $\varepsilon_d/U < -1/2$, which 
increases the value of the thermopower.
However, if the temperature 
 is larger than the {\it SU}(4) Kondo temperature,
the Kondo resonance is smeared by thermal fluctuations, and therefore the thermopower has a minimum in the  regime $-1/2 <\varepsilon_d/U <0$.
As the energy level is further lowered,
the {\it SU}(4) Kondo effect and the resulting thermopower are both suppressed.
Note that the Hund coupling hardly affects either the thermopower or conductance, as shown in Fig. \ref{fig:double_vjSG}(a) and (b), because of $n_d \sim 1$ in this regime ({\it i.e.}
 configurations with single-electron occupation are dominant).
%%%%%%%%%%%%%%%%%%%%%
%%%%%%%%%%%%%%%%%%%%%
Since the region of $-3<\varepsilon_d/U < -2$ can be related to $-1<\varepsilon_d/U < 0$ via an electron-hole transformation,
we can directly apply the above discussions on the {\it SU}(4) Kondo effect to the former region by changing the sign of the thermopower.

Let us now turn to the region of $-2<\varepsilon_d/U<-1$, where $n_d \sim 2$.
At $J=0$, the Kondo effect due to six-fold degenerate electron 
configurations in the QD occurs.
Although such an orbital Kondo effect indeed emerges around $\varepsilon_d/U=-3/2$ in this case,  the thermopower is almost zero there because the Kondo resonance is located just at the Fermi level.
Therefore, 
%%when the dot level is changed, the position of the Kondo resonance
%%is shifted  across the Fermi level, which causes the 
%%sign change of the thermo power.
around $\varepsilon_d/U=-3/2$, even small perturbations could easily change the sign of the thermopower at low temperatures.
Note that these properties are quite similar to those for the ordinary spin Kondo effect shown in Fig. \ref{fig:sorbtp}(a), because  the filling is almost half in both cases.
We next discuss the influence of the Hund coupling $J$ in this region.
For large Hund couplings $J$, the triplet Kondo effect is realized and the resulting Kondo temperature becomes very small,
so that the thermopower shown in Fig. \ref{fig:double_vjSG}(a)
is dramatically suppressed by thermal fluctuations.
Such suppression of the Kondo effect by the Hund coupling is also
 observed in the conductance shown in Fig. \ref{fig:double_vjSG}(b).

\begin{figure*}[tb]
\begin{center}
\includegraphics[width=0.8\linewidth]{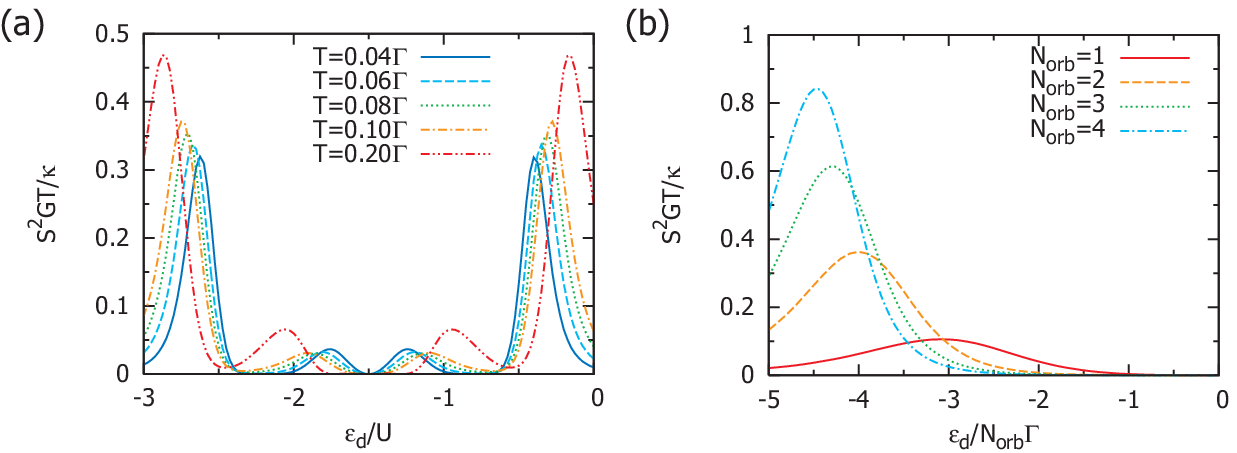}
\end{center}
\caption{(Color online) 
(a)Figure of merit $ZT=S^2GT/\kappa$ due to the two-orbital Kondo effect ($N_{orb}=2$), as a function of the QD level.
We set $J=0$ and $U=8\Gamma$.
(b)Comparison of the figure of merit due to the Kondo effect with several choices of orbital degrees of freedom $N_{orb}$, as a function of QD levels, for $U=\infty$ and $T=0.01\Gamma$.
}
\label{fig:double_vFOM}
\end{figure*}

Lastly, the thermoelectric figure of merit as a function of the QD level 
is shown in Fig. \ref{fig:double_vFOM}.
%%%%%%%%%%%%%%%%%%%%%
%%%%%%%%%%%%%%%%%%%%%
It turns out that $ZT$ is still smaller than unity in the whole
 parameter region shown in Fig. \ref{fig:double_vFOM}.
However, it is seen that in the {\it SU}(4) Kondo region ($-1 <\varepsilon_d/U<0,-3 <\varepsilon_d/U<2$), $ZT$ has a larger value than that of the spin Kondo effect discussed in the previous section.
In general, larger orbital degeneracy yields larger figure of merit as shown in Fig. \ref{fig:double_vFOM} (b).
Therefore, we expect that if the control parameters are properly
tuned for highly symmetric QD systems, the figure of merit can be improved
substantially.

%%%%%%%%%%%%%%%%%%%%%%%%%%%%%%%%%%%%%%%%%
\subsubsection{Magnetic-filed control}%%%
\begin{figure*}[bt]
%h=here, t=top, b=bottom, p=separate figure page
\begin{center}\leavevmode
\includegraphics[width=0.8\linewidth]{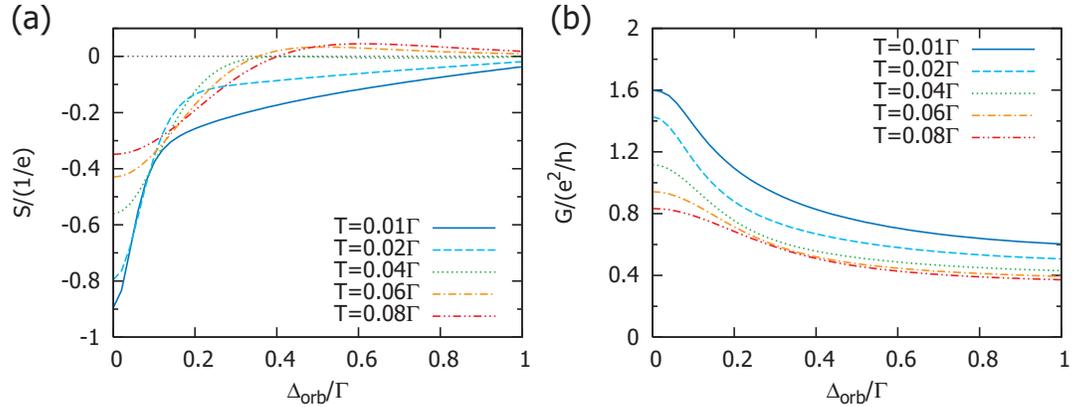}
\caption{(Color online)
(a) Thermopower and (b) conductance for the two-orbital QD system, 
in case of
$\varepsilon_d=-U/2$, as a function of the orbital splitting $\Delta_{orb}$.
We set $U=8\Gamma$.
 }
\label{fig:double_kSG}
\end{center}
\end{figure*}

%%%%%%%%%%%%%%%%%%%%%%%%%%%%%%%%%%%%%%%%%
Let us now analyze the effects of orbital splitting caused by magnetic fields.
The computed thermopower for  $\varepsilon_d/U=-1/2$ is shown in Fig. \ref{fig:double_kSG} as a function of the orbital splitting $\Delta_{orb}$.
%%%%%%%%%%%%%%%%%%%%
%%%%%%%%%%%%%%%%%%%%%
It is seen that magnetic fields dramatically suppress the thermopower, in contrast to the monotonic and moderate decrease of the conductance. The suppression 
is caused by the following two distinct mechanisms.
In the presence of magnetic fields, the Kondo effect gradually changes 
from the {\it SU}(4) orbital type to the {\it SU}(2) spin type because 
the orbital degeneracy is lifted.
As a consequence, the effective Kondo temperature (resonance width)
is reduced substantially, so that the thermopower at higher temperatures is easily reduced in the presence of magnetic fields (e.g. $T \sim 0.08 \Gamma$, $0.06 \Gamma$).
On the other hand, at low temperatures close to the {\it SU}(2) Kondo temperature $T \sim T_K^{SU(2)} \sim 0.01\Gamma$ ($T_K^{SU(2)}$ is estimated from the temperature dependence of conductance shown in Fig. \ref{fig:sorbtd}(b)), the renormalized Kondo resonance itself is well developed, but its peak position approaches the Fermi level, giving rise to the strong reduction of thermopower.
%%These effects are also confirmed the conductance behavior 
%%shown in Fig. \ref{fig:double_kSG} (b).
In strong fields, the effective Kondo resonance is located  very closely to 
the Fermi level, so that the thermopower can even change its sign, which is 
indeed seen in Fig.  \ref{fig:double_kSG} (a)

Finally a brief comment is in order for other choices of the parameters.
The thermopower for $\varepsilon_d/U=-5/2$ shows similar magnetic-field dependence to the $\varepsilon_d/U=-1/2$ case except that its sign is changed.
For $\varepsilon_d/U=-3/2$, the thermopower is almost zero and independent of magnetic fields,
because the Kondo resonance is pinned at the Fermi level and gradually disappears with increase of magnetic fields.

%% file: Sec/summary.tex
%%%%%%%%%%%%%%%%%%%%%%%%%%%%%%%%%%%%%%%
\section{Summary}\label{sec:summary}%%%
%%%%%%%%%%%%%%%%%%%%%%%%%%%%%%%%%%%%%%%
We have studied the thermopower due to the spin and orbital Kondo effect in a single quantum dot system connected to two normal leads under gate-voltage and magnetic-field control.
In particular, making use of the NCA method for the Anderson model with finite Coulomb repulsion,
we have systematically investigated the low-temperature properties in several distinct electron-charge regions.
It has been elucidated how the asymmetric nature of the resonance due to the orbital Kondo effect controls the magnitude and the sign of the thermopower at low temperatures.
For $\varepsilon_d/U \sim -1/2 (\varepsilon_d/U \sim -5/2)$, where $n_d \sim 1(3)$,
the SU(4) Kondo effect is dominant and the corresponding 
thermopower is enhanced.
%%These two regions are related to each other via an electron-hole 
%%transformation, which merely gives an opposite sign of the thermopower.
In addition, magnetic fields change the Kondo effect from {\it SU}(4) to an 
{\it SU}(2) type, resulting in two major effects:
the effective resonance position approaches the Fermi level and the Kondo temperature is decreased, which reduces the thermopower in magnetic fields.
On the othr hand, it hase been found that 
gor $\varepsilon_d/U \sim -3/2$, where $n_d \sim 2$, the Kondo effect due to six-fold degenerate states occurs for $J = 0$.
However, the thermopower is strongly reduced  in this case because the resonance peak is located near the Fermi level. The introduction of the
 Hund coupling makes the triplet Kondo effect dominant, so that
the resulting small Kondo temperature futher suppresses the thermopower 
around $\varepsilon_d/U \sim -3/2$ at finite temperatures.
%%In this region, magnetic fields do not affect the asymmetry 
%%of the resonance peak and the resulting thermopower remains 
%%almost zero because the filling is fixed. 

As summarized above, the thermopower due to the orbital Kondo effect exhibits a variety of properties, which reflect the detailed shape of the effective Kondo resonance near the Fermi level. In particular, it sensitively depends on the orbital degeneracy, the strength of Coulomb/Hund interactions, the magnetic field, etc. In this way, the thermopower can be used to explore the detailed nature of local electron correlations. For the purpose of  more practical use, we have  addressed the application of QD systems to some thermoelectric devices. Judging from the figure of merit obtained,  such applications may be still limited. Nevertheless,  we have demonstrated that if we can use the orbital degeneracy appropriately,  the figure of merit can be improved substantially.

%% file: Sec/acknowledgment.tex
%%%%%%%%%%%%%%%%%%%%%%%%%%%%%
\section*{Acknowledgment}%%%%
%%%%%%%%%%%%%%%%%%%%%%%%%%%%%
We thank S. Tarucha, A. C. Hewson, M. Eto, A. Oguri, S. Amaha and T. Ohashi for  valuable discussions.
RS was supported by the Japan Society for the Promotion of Science.